\documentclass[aps,prb,a4paper,twocolumn,superscriptaddress,floatfix,showpacs]{revtex4-1}
\usepackage{amsmath}
\usepackage{amssymb}
\usepackage{graphicx}
\usepackage{color}
\usepackage{epstopdf}
\usepackage{subfigure}
\begin{document}

\title{Nonmonotonous classical magneto-conductivity of a two-dimensional electron 
gas in a disordered array of obstacles}
\author{N. H. Siboni}
\affiliation{Institut f\"ur Theoretische Physik II, Heinrich-Heine-Universit\"at, Universit\"atsstra\ss e 1, 40225 D\"usseldorf, Germany}
\author{J. Schluck}
\affiliation{Institut f\"ur Experimentelle Physik der kondensierten Materie, Heinrich-Heine-Universit\"at, Universit\"atsstra\ss e 1, 40225 D\"usseldorf, Germany}
\author{K. Pierz}
\affiliation{Physikalisch-Technische Bundesanstalt, Bundesallee 100, 38116 Braunschweig, Germany}
\author{H. W. Schumacher}
\affiliation{Physikalisch-Technische Bundesanstalt, Bundesallee 100, 38116 Braunschweig, Germany}
\author{D. Kazazis}
\affiliation{CNRS, Univ. Paris-Sud, Universit\'{e} Paris-Saclay, C2N Marcoussis, 91460 Marcoussis, France}
\altaffiliation [Present address: ]{Laboratory for Micro- and Nanotechnology, Paul Scherrer Institute, 5232 Villigen-PSI, Switzerland.}
\author{J. Horbach}\email{horbach@thphy.uni-duesseldorf.de}
\affiliation{Institut f\"ur Theoretische Physik II, Heinrich-Heine-Universit\"at, Universit\"atsstra\ss e 1, 40225 D\"usseldorf, Germany}
\author{T. Heinzel}\email{Thomas.Heinzel@hhu.de}
\affiliation{Institut f\"ur Experimentelle Physik der kondensierten Materie, Heinrich-Heine-Universit\"at, Universit\"atsstra\ss e 1, 40225 D\"usseldorf, Germany}
\date{\today}

\begin{abstract}
Magnetotransport measurements in combination with molecular dynamics (MD)
simulations on two-dimensional disordered Lorentz gases in the classical
regime are reported. In quantitative agreement between experiment and
simulation, the magnetoconductivity displays a pronounced peak as a
function of perpendicular magnetic field $B$ which cannot be explained
in the framework of existing kinetic theories. We show that this peak
is linked to the onset of a directed motion of the electrons along the
contour of the disordered obstacle matrix when the cyclotron radius
becomes smaller than the size of the obstacles. This directed motion
leads to transient superdiffusive motion and strong scaling corrections in
the vicinity of the insulator-to-conductor transitions of the Lorentz gas.
\end{abstract}
\maketitle

A system of non-interacting particles moving in a Poisson
distributed array of identical obstacles is known as a Lorentz
gas. Originally proposed for the motion of electrons in a metal
\cite{Lorentz1905}, the Lorentz gas has developed into a universal
model for transport phenomena in many types of heterogeneous
media, like anomalous diffusion in colloidal and bio-systems
\cite{Hofling2006,Horton2010,Hofling2013,Skinner2013,Schnyder2015,Zeitz2017},
microwave-induced magnetoresistance oscillations \cite{Beltukov2016},
or negative magnetoresistance in metallic and semiconductor systems
\cite{Altshuler1980,Mirlin2001a,Polyakov2001,Dmitriev2002,Dmitriev2012,Bockhorn2014}.
Versatile implementations of Lorentz gases can be realized experimentally
by two-dimensional electron gases (2DEGs) exposed to a random array of
obstacles. Such systems provide a high intrinsic electron mobility and
the option to pattern the obstacles lithographically.  A perpendicular
magnetic field $B$ tunes the cyclotron radius $R_{\rm cy}\propto B^{-1}$
of the electronic motion, acting as an additional characteristic length
scale.

Experimental studies of the magnetotransport of 2DEGs
in disordered obstacle arrays have been scarce (see, e.g.,
Refs.~\cite{Gusev1994,Tsukagoshi1995,Nachtwei1997,Nachtwei1998,Yevtuchenko2000}).
Especially, the magnetoconductivity $\sigma_{xx}(B)$ in a regime of
high obstacle densities $n^\star$ and large magnetic field has not
been systematically explored up to now. Here, $n^\star$ denotes the
dimensionless number density of obstacles, $n^\star = \frac{N}{A} R_{\rm
int}^2$, with $N$ the number of obstacles, $A$ the area of the system
and $R_{\rm int}$ the interaction distance between an electron and an
obstacle, i.e., the effective radius of the (circular) obstacles. In
this letter, we present a semi-classical experimental realization of
a Lorentz gas in combination with classical molecular dynamics (MD)
simulations and demonstrate that the electron transport qualitatively
changes if the cyclotron radius $R_{\rm cy}$ becomes smaller than the
interaction distance $R_{\rm int}$. As expected from kinetic theories,
the conductivity $\sigma_{xx}(B)$ is a monotonously decaying function
at low densities. For large densities and $R_{\rm cy}\lesssim R_{\rm
int}$, however, it exhibits a maximum that moves to larger values of
$B$ with increasing $n^\star$.  This maximum has been observed in
simulations \cite{Kuzmany1998,Schirmacher2015} but it has hitherto
remained unexplained and not been observed experimentally.

The magneto-transport in the 2D Lorentz gas is associated with two
insulator-to-conductor transitions at high and low number density
$n^\star$ of the obstacles which are due to underlying static percolation
transitions \cite{Kuzmany1998,Schirmacher2015}. The location of the
transition at high density is independent of the magnetic field
$B$ and located at a critical density $n_c^{\star}= 0.359$ for
a Poisson-distributed arrangement of overlapping disks: while for
$n<n_c^{\star}$ the electron exhibits diffusive transport through the
void space between the obstacles, for $n>n_c^{\star}$ the void space is
disconnected into finite pockets in which the electron is trapped. The
second, $B$-dependent localization transition occurs at a density
$n^\star_{\rm ld, c}(B)< n_c^\star$. It can be understood in terms of
skipping orbits that the electrons, acting as tracer particles in this
experimental implementation, perform around the obstacles, or clusters
thereof, which localizes all particles as $n^\star$, or the cyclotron
radius $R_{\rm cy}$, respectively, is decreased \cite{Kuzmany1998}.

For a fixed $B$ field, the magnetoconductivity exhibits a maximum as
a function of $n^\star$ which is located at $n_{\rm ld, c}^\star <
n_{\rm max}^\star < n_c^\star$. This maximum is intimately related to
the maximum in $\sigma_{xx}(B)$. As elucidated by our MD simulations,
the line of maxima, $n_{\rm max}^\star(B)$, follows a simple law (see
below) and merges with the two critical points in the limit $B\to
\infty$. The change of the transport for $R_{\rm cy} \lesssim R_{\rm
int}$ along the line $n_{\rm max}^\star(B)$ is due to a change of the
motion of the tracer particles (electrons) from a diffuse scattering
by the obstacles to a directed motion along the contour of the obstacle
arrangement. In the limit $B\to \infty$, this directed motion completely
dominates the transport and suppresses the critical slowing down at
$n_{\rm ld, c}^\star$ and $n_c^\star$. At finite but high $B$ fields it
leads to strong corrections to the scaling behavior in the vicinity of
the critical points.

\begin{figure}
\includegraphics[scale=0.9]{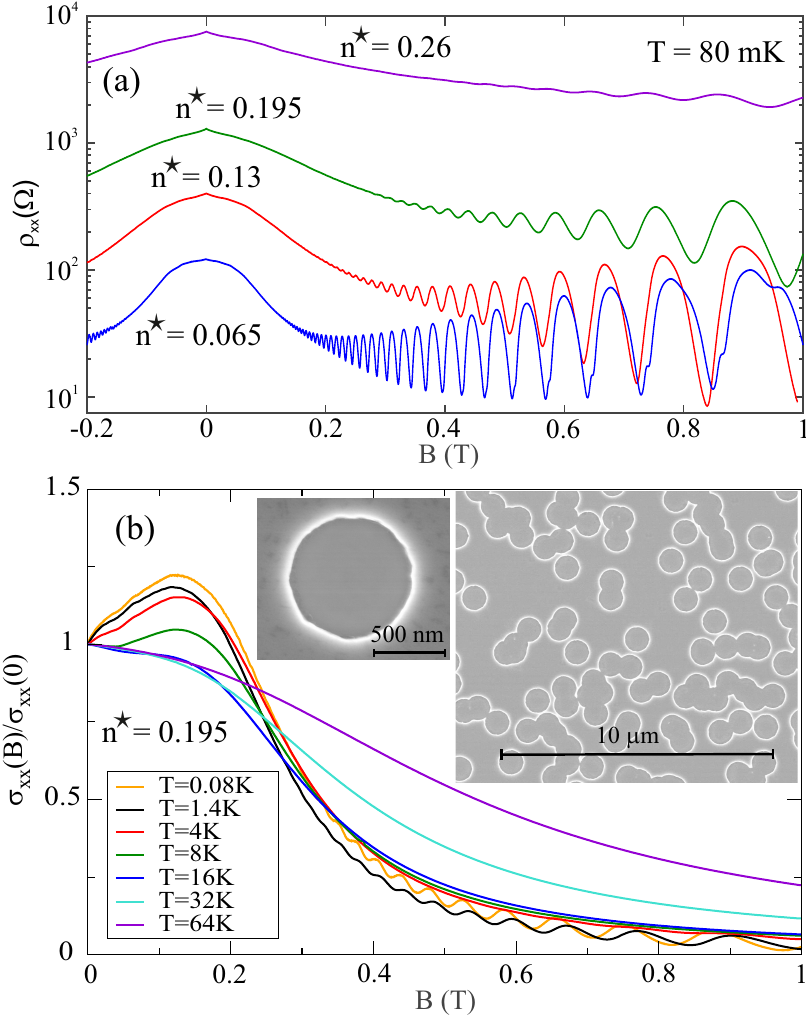}
\caption{\label{fig1} (a) The longitudinal magnetoresistivities
$\rho_{xx}$ of the Lorentz gases of various obstacle densities $n^\star$,
measured at a temperature of $\approx 80 \,\mathrm{mK}$. (b) The
magnetoconductivity $\sigma_{xx} (B)$ and its temperature dependence, as
determined from the measurements for the $n^\star=0.195$ array. Insets:
Scanning electron microscope picture of a Lorentz array section with
$n^\star=0.195$ and magnified view of a single disk.}
\end{figure}
{\it Experiment.} A $\mathrm{GaAs/Al_{0.3}Ga_{0.7}As}$ heterostructure
with a 2DEG located $150 \,\mathrm{nm}$ below the surface is used. The
electron density and mobility are $n_e=2.5\times10^{15}\,\mathrm{m^{-2}}$
and $340\,\mathrm{m^2/Vs}$, respectively, corresponding to a mean free
path of $31\,\mathrm{\mu m}$ at temperatures below $1\,\mathrm{K}$. The
repulsive Lorentz obstacles are formed by circular holes in the 2DEG. They
are patterned by electron beam lithography and subsequent inductively
coupled reactive ion etching. Within each array, the obstacles are
nominally identical in shape and size, while their positions are Poisson
distributed, with mutual overlaps allowed. All disks have a lithographic
radius of $425\,\mathrm{nm}$, see Fig.~\ref{fig1}(a). From Aharonov-Bohm
measurements in large magnetic fields \cite{Iye2004}, we estimate the
lateral depletion length to $\approx 75\,\mathrm{nm}$ \cite{Schluck2015},
such that the effective electronic disk radius is $R_{int}\approx
500\,\mathrm{nm}$. Besides a Hall bar without intentional obstacles, the
chip contains four Lorentz arrays with disk densities $n^\star=0.065$,
$0.13$, $0.195$, and $0.26$, respectively. The arrays have an area of
$200\,\mathrm{\mu m}$ by $100\,\mathrm{\mu m}$. The mean free path due to
the scattering at the disks is $4\,\mathrm{\mu m}$, $2\,\mathrm{\mu m}$,
$1.3\,\mathrm{\mu m}$, and $1\,\mathrm{\mu m}$, respectively.

The samples were inserted in the mixing chamber of a dilution refrigerator
with a base temperature of $8\,\mathrm{mK}$. The electron temperature
is estimated to $\approx 80\,\mathrm{mK}$. A $\mathrm{^4He}$ gas flow
cryostat with a variable temperature insert and a base temperature
of $1.4\,\mathrm{K}$ is used for measurements at temperatures above
$1\,\mathrm{K}$. An AC current ($500\,\mathrm{nA}$, $17.7\,\mathrm{Hz}$)
is injected. The longitudinal and Hall voltages are measured at suitable
probes using lock-in amplifiers.

The longitudinal magnetoresistivity $\rho_{xx}(B)$ (see
Fig.~\ref{fig1}(a)) shows a strong peak around $B=0$ which in some
arrays extends well into the range where Shubnikov--de Haas oscillations
\cite{Ando1982} are observed. As $n^\star$ is increased from $0.065$
to $0.26$, $\rho_{xx}(0)$ increases by approximately a factor of 50.

The longitudinal magnetoconductivity is obtained
from the measured resistivity components via
$\sigma_{xx}(B)=\rho_{xx}(B)/(\rho^2_{xx}(B)+\rho^2_{xy}(B))$, where
$\rho_{xy}(B)$ denotes the Hall resistivity (see the supplement for the
corresponding measurements). In Fig.~\ref{fig1}(b), the thereby obtained
$\sigma_{xx}(B)$ is shown for the array with $n^\star=0.195$ for various
temperatures. A pronounced maximum at $B\approx 140\,\mathrm{mT}$
is observed. It shows a weak temperature dependence and evolves
at higher temperatures into a shoulder that is still visible at
$32\,\mathrm{K}$. This weak temperature dependence indicates a classical
origin. We have observed the same phenomenology in a set of scaled samples
with identical number densities but with $R_{int}=1\,\mathrm{\mu m}$
(not shown). This behavior is in qualitative contradiction to both the
Boltzmann model as well as to the Bobylev model valid for Lorentz gases
with small $n^\star$ \cite{Bobylev1995}. Rather, it is associated with the
above mentioned conductivity maximum as predicted by numerical simulations
for high density Lorentz gases \cite{Kuzmany1998,Schirmacher2015}.

{\it Simulations.} Classical molecular dynamics of a system of
non-interacting fluid particles in a two-dimensional matrix of randomly
placed obstacle particles are performed using LAMMPS \cite{LAMMPS}
with a modified integrator to include the magnetic field. Matrix (index
M) and fluid particles (F) interact via a shifted, purely repulsive
Weeks-Chandler-Andersen (WCA) potential, $u_{\rm FM}(r) = 4 \epsilon
\left[ (R_{\rm int}/r)^{12} - (R_{\rm int}/r)^6 + 1/4 \right]$ for
$r < 2^{1/6}R_{\rm int}$ and $u_{\rm FM}(r) = 0$ otherwise. Here, we
have set the energy parameter to $\varepsilon=0.1\,\varepsilon_{\rm
M}$ and the interaction range to $R_{\rm int}=0.5\,\sigma_{\rm
M}$ where $\varepsilon_{\rm M}=1.0$ and $\sigma_{\rm M}=1.0$
correspond to the energy parameter and the diameter of the obstacle
particles, respectively. For the comparison between simulation and
experiment exactly the same configurations of obstacles as in the
experiment are implemented. For the other calculations, we use 100
statistically independent matrix structures at each number density,
$n^{\star}=\frac{N}{L^2} R_{\rm int}^2$ with $N$ the number of matrix
particles and $L$ the linear dimension of the simulation square.

Newton's equations of motion are integrated numerically using
the velocity-Verlet algorithm \cite{Binder2004} with a time step of
$10^{-3}t_0$ with $t_0 := [m(\sigma_{\rm M})^2/\varepsilon_{\rm M}]^{1/2}$
and $m=1.0$ the mass of a fluid particle. The particles carry a charge
$e=1$ and a mass $m=1$, and are subjected to a uniform magnetic field
$B$ that acts perpendicular to the plane of motion. The velocity of the
fluid particles is fixed to a constant magnitude $v_{\rm F}=\sqrt{2}$,
associated with a cyclotron radius of $R_{\rm cy}=mv_{\rm F}/(eB)$
or $R_{\rm cy} = R_{\rm int}/\tilde{B}$, with $\tilde{B}$ being the
dimensionless magnetic field $\tilde{B} = B/B_0$ (with $B_0=\frac{mv_{\rm
F}}{e R_{\rm int}}$). Between 100 and 2400 fluid particles per host
structure are used for runs of up to $10^6 t_0$. For the calculation of
time averages, 10 time origins per run are used, spaced equidistantly
over the whole simulation time.

The conversion of units between simulation and experiment is
as follows: $\sigma_{\rm M}=10^{-6}$\,m (obstacle diameter),
$m=6.097\times10^{-32}$\,kg (effective electron mass in GaAs), $t_0=9.226\times
10^{-12}$\,s, $e=1.6\times 10^{-19}$\,C (electron charge),
and $B_0 = 0.168$\,T.

\begin{figure}
\subfigure{\includegraphics[width=0.49\columnwidth]{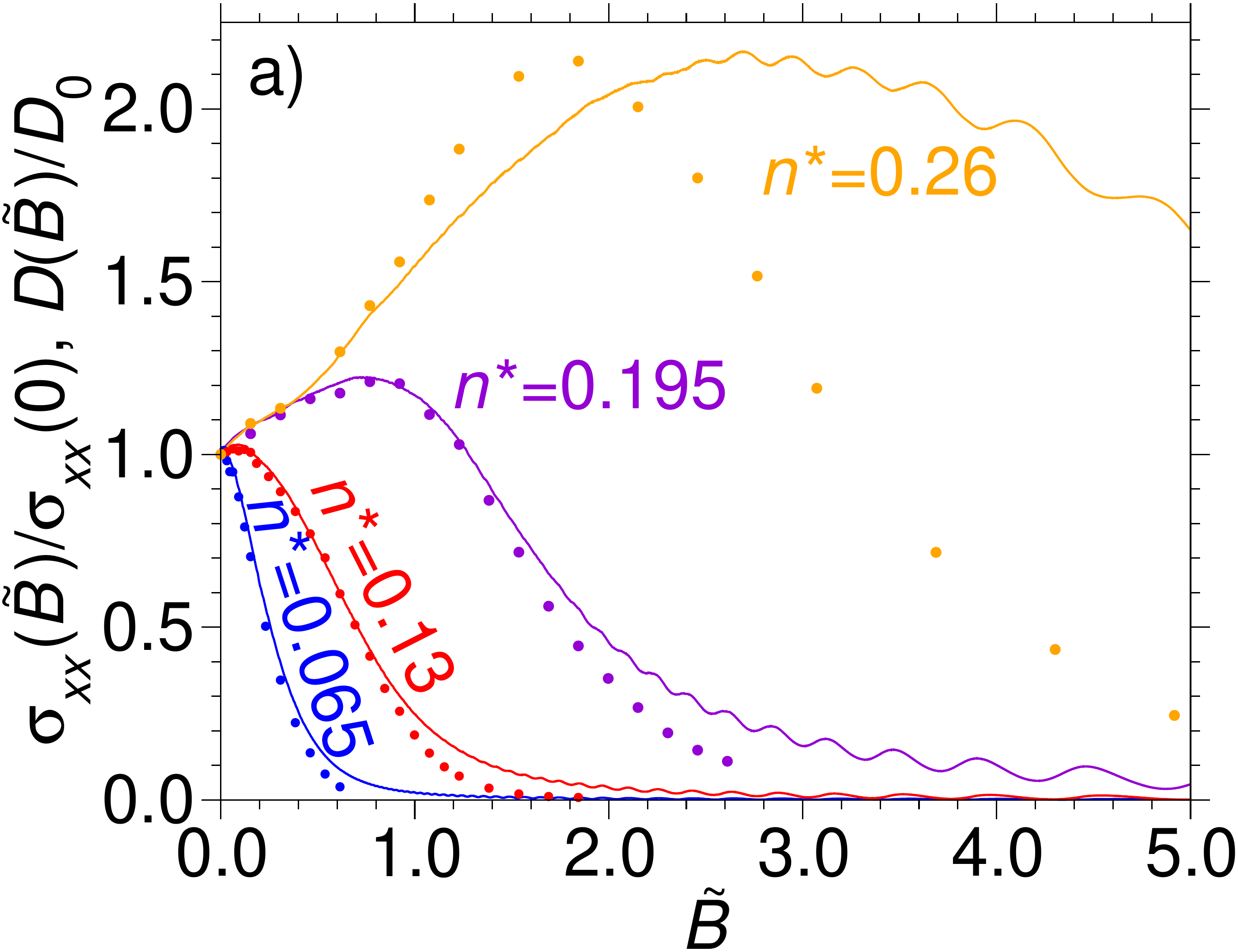}}
\subfigure{\includegraphics[width=0.49\columnwidth]{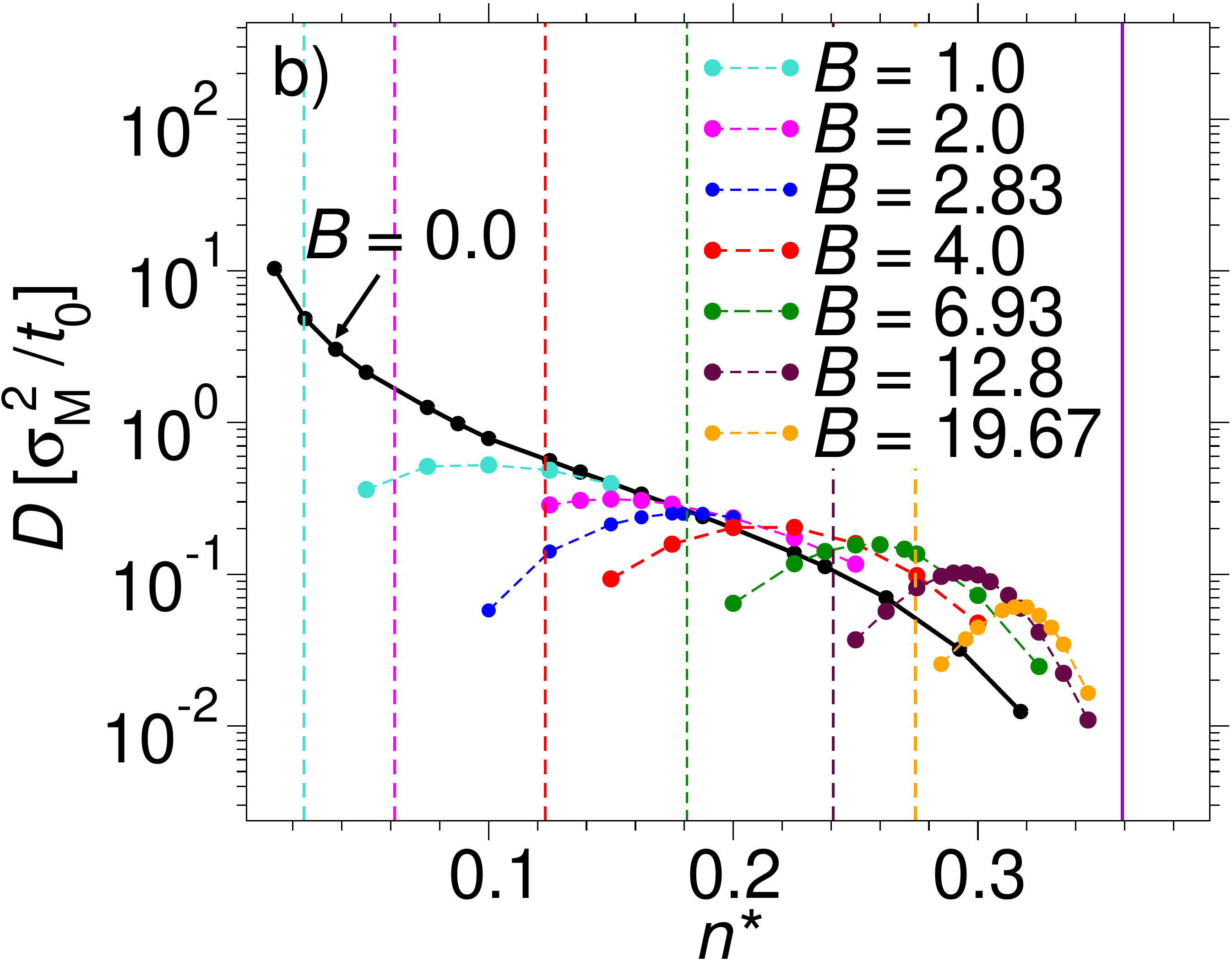}}
\includegraphics[angle=90,scale=0.13]{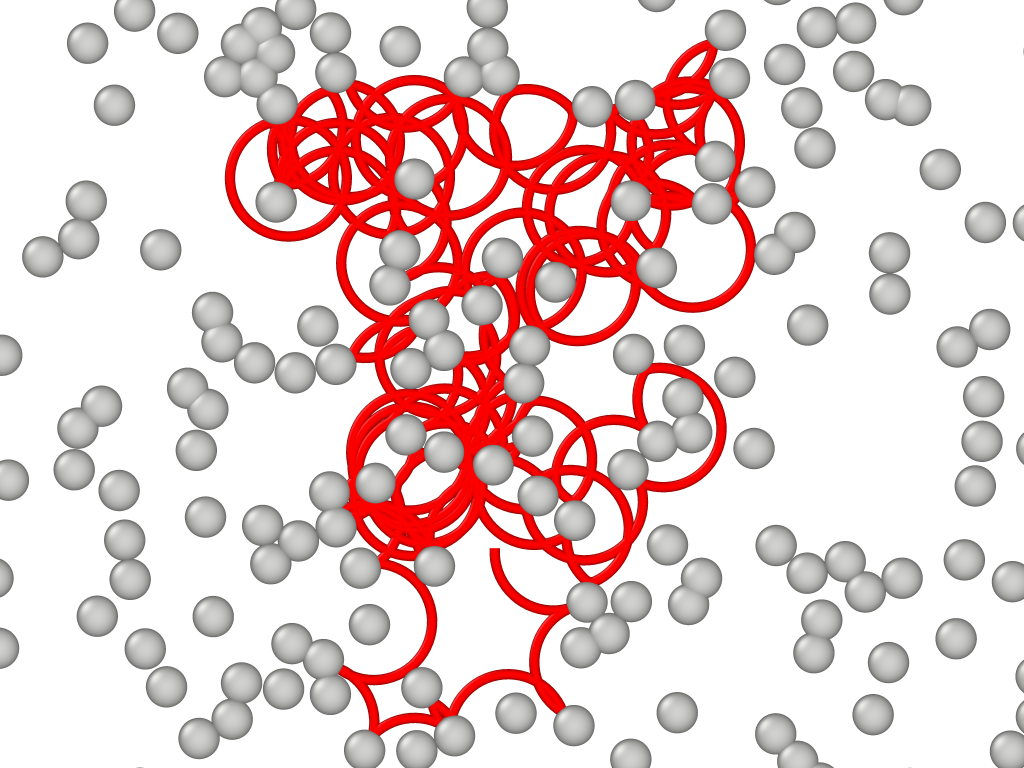}
\includegraphics[angle=90,scale=0.13]{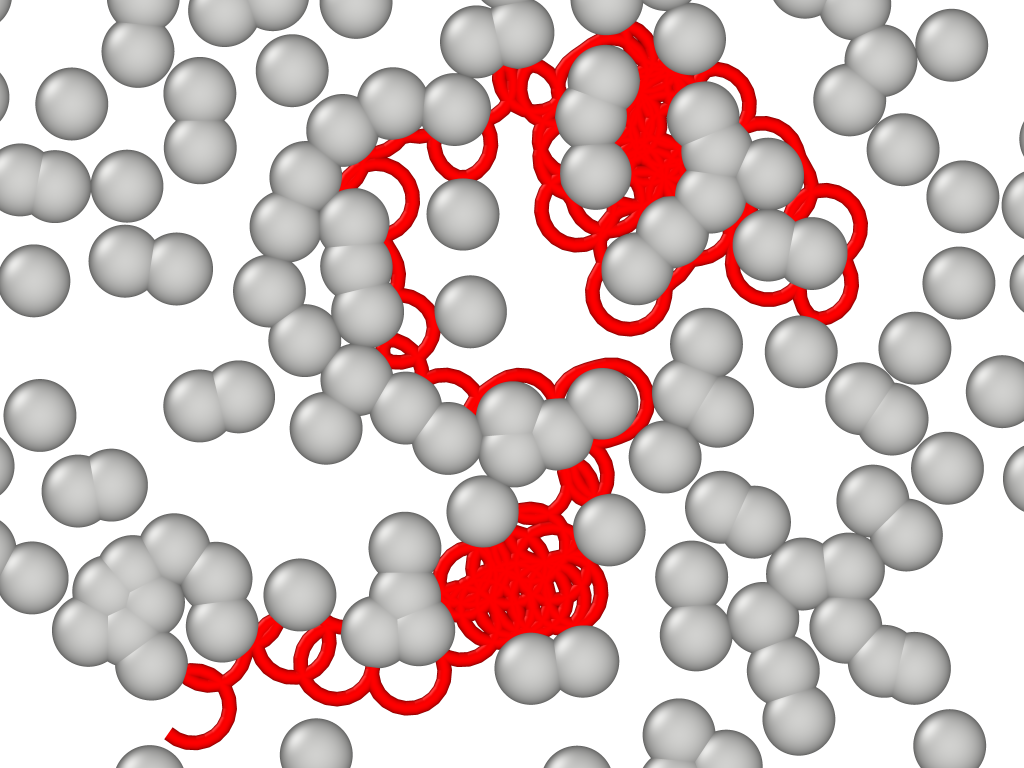}
\includegraphics[angle=90,scale=0.21]{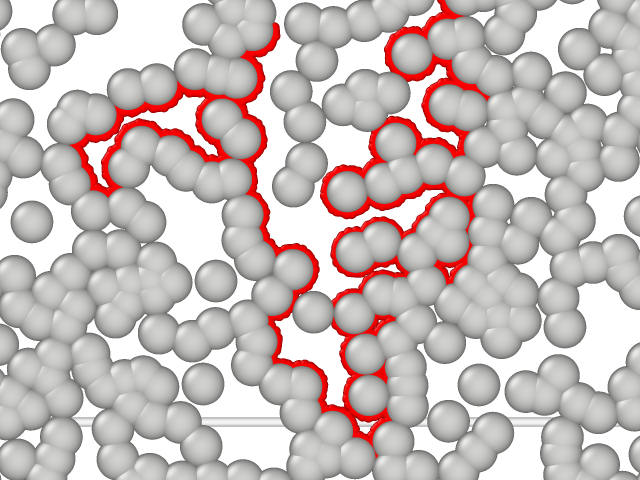}
\caption{\label{fig2} (a) $\sigma_{xx}(\tilde{B})/\sigma_{xx}(0)$ from 
experiment (full lines) in comparison to $D(\tilde{B})/D(0)$ from
simulation (full circles) for different values of $n^\star$. 
(b) $D(n^\star/n^\star_{\rm c})$ from the simulation for different 
values of $\tilde{B}$. The vertical lines correspond to the locations of 
$n^\star_{\rm c}$ (solid line) and $n^\star_{\rm ld, c}(\tilde{B})$ 
(dashed lines). The snapshots correspond to the location of the maxima 
in $D(n^\star)$ for $\tilde{B}=0.35$, 1.41, and 6.95 (from left to right).
The trajectories are illustrated using OVITO \cite{Ovito}.}
\end{figure}
{\it Results.} For a system of non-interacting charged particles
(electrons), the conductivity $\sigma_{xx}$ is directly related
to the self-diffusion constant $D$ via $\sigma_{xx}=\frac{n_e
e^2}{m} D$.  Hence, we can directly compare the experimentally
obtained conductivity, normalized to its value at $\tilde{B}=0$,
$\sigma_{xx}(\tilde{B})/\sigma_{xx}(0)$, to the corresponding ratio of
diffusion constants from the MD simulation, $D(\tilde{B})/D(0)$.  In the
simulation, the self-diffusion constant can be obtained from the long-time
limit of the mean-squared displacement (MSD) of a tagged particle,
$\delta r^2(t)$, using the Einstein relation $D = \lim_{t\to \infty}
\delta r^2(t)/4t$. Here, the MSD is defined as $\delta r^2(t) = \langle
(\vec{r}(t) - \vec{r}(0))^2 \rangle$, with $\vec{r}(t)$ the position of
the particle at time $t$ and $\langle \cdots \rangle$ an ensemble average.

The comparison of $\sigma_{xx}(\tilde{B})/\sigma_{xx}(0)$ from the
experiment with $D(\tilde{B})/D(0)$ from the simulation is shown in
Fig.~\ref{fig2}a for different number densities. At low $\tilde{B}$
fields, simulation and experiment are in good agreement. We note that
therefore, both weak localization corrections \cite{Altshuler1982} and
interaction effects \cite{Alekseev2016} can be excluded as possible
origin. The experimental values are significantly larger than the
ones from the simulation at high $\tilde{B}$ fields. We tentatively
attribute these deviations to a combination of quantum effects like
the onset of Shubnikov--de Haas oscillations, and depinning of
electrons from the obstacle clusters by the residual random disorder
\cite{Mirlin2001a}. However, different from the monotonous decay of
$\sigma_{xx}(\tilde{B})/\sigma_{xx}(0)$ and $D(\tilde{B})/D(0)$ at low
densities, a maximum occurs at large densities (see, e.g.~the results
for $n^\star = 0.195$ where simulation and experiment
are in very good agreement around the maximum).

To get further insight into the nature of this change we plot in
Fig.~\ref{fig2}(b) the diffusion constant from the simulation as function
of number density, $D(n^\star)$, for different values of $\tilde{B}$,
including $\tilde{B}=0$. At a given finite value of $\tilde{B}$, the
diffusion coefficients vanish at the critical densities $n^\star_{\rm c}$
and $n^\star_{\rm ld, c}(\tilde{B})<n^\star_{\rm c}$.  Therefore, at a
given value of $\tilde{B}$, the function $D(n^\star)$ exhibits a maximum
in the interval $[n^\star_{\rm ld, c}(B),n^\star_{\rm c}]$. The snapshots
in Fig.~\ref{fig2} show typical trajectories for different densities
corresponding to the maxima in $D(n^\star)$ at $\tilde{B}=0.35$, 1.41
and 6.95 from left to right (cf.~corresponding movies in the supplementary
material). These snapshots indicate a qualitative
change of the motion around $\tilde{B}=1$, from a diffuse scatter of
the tracer particle by the obstacle for $\tilde{B}\ll 1$ to a directed
motion along the contour of the obstacle network for $\tilde{B}\gg 1$. The
trajectory at $\tilde{B}=1.41$ indicate a mixture of diffuse scattering
and directed motion.

\begin{figure}
\includegraphics[scale=0.3]{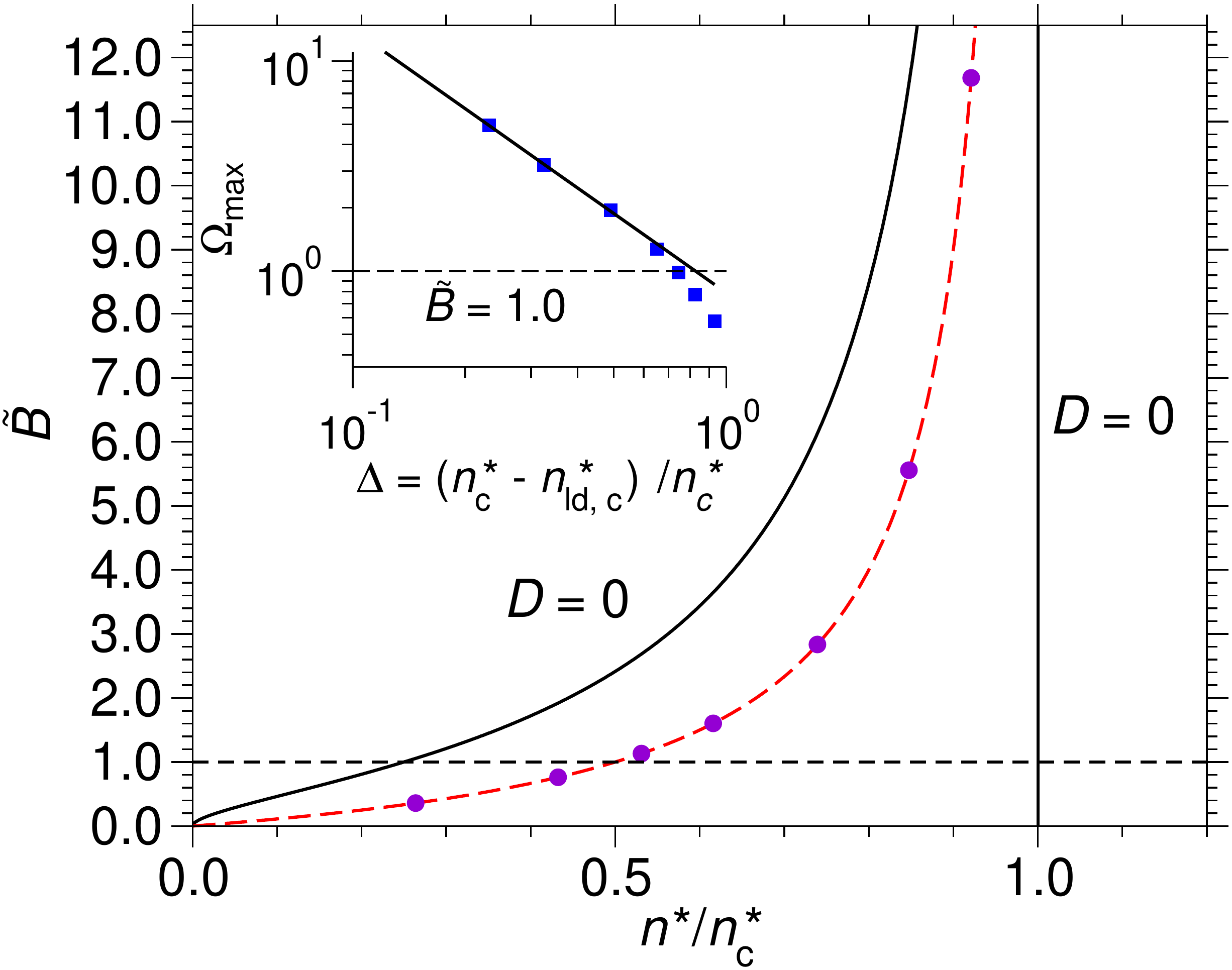}
\caption{\label{fig3} Phase diagram in terms of $\tilde{B}$
vs.~$n^\star/n^\star_{\rm c}$. The red dashed line is obtained from
Eq.~(\ref{eq_bmax}) and filled blue circles show $\tilde{B}_{\rm max}$,
as determined from the simulation (cf.~Fig.~\ref{fig2} (b)). The inset shows
$\Omega_{\rm max}$ as function of $\Delta$, with the solid line
being a fit with $\Omega_{\rm max} \propto \Delta^{-1.25}$ (see text).}
\end{figure}
The phase diagram in Fig.~\ref{fig3} shows $\tilde{B}$
vs.~$n^\star/n^\star_{\rm c}$, with the two lines of critical points
at low and high density. While the critical points at high density
are independent of $\tilde{B}$ at $n^\star/n^\star_{\rm c}=1$,
the low density critical points are located at $\tilde{B}_{\rm
ld, c} = \left(\sqrt{n^\star_{\rm c}/n^\star} - 1\right)^{-1}$
\cite{Kuzmany1998}. The dashed red line in between the two critical lines
in Fig.~\ref{fig3} corresponds to the points of maximal diffusion. Its
form can be understood as follows: The density $n_{\rm max}(\tilde{B})$ at
which the diffusion coefficient is maximal is associated with two limiting
cases. The maximum vanishes towards $\tilde{B}\to 0$, i.e.~$n_{\rm
max}(\tilde{B}\to 0)= 0$, and it should coincide with $n^\star_{\rm c}$
in the limit $\tilde{B}\to \infty$ (then $R_{\rm cy}=0$ and $n^\star_{\rm
c}=n^\star_{\rm ld, c}$). A function that interpolates between these two
limiting cases is $n_{\rm max}^\star = n_{\rm c}^\star R_{\rm int}\,
(R_{\rm int} + R_{\rm cy})^{-1}$. When this expression is solved for
$\tilde{B}$, one obtains the following law for the density dependence
of the reduced magnetic field at maximal diffusion:
\begin{equation}
\label{eq_bmax}
\tilde{B}_{\rm max}=\left( n^\star_{\rm c}/n^\star - 1 \right)^{-1}
\end{equation}
The points that are on the maximal diffusion curve in Fig.~\ref{fig3}
are directly obtained from the data in Fig.~\ref{fig2}b, confirming
that Eq.~(\ref{eq_bmax}) indeed holds. Along the line of maxima, the
transport of the tracer particle changes around $\tilde{B}=1$. This
can be inferred from the inset in Fig.~\ref{fig3} where the ratio of
the diffusion constant at the maximum to that at $\tilde{B}=0$ at the
corresponding density, $\Omega_{\rm max}=D_{\rm max}(B)/D(n^\star_{\rm
max}, \tilde{B}=0)$, is plotted as a function of the distance
between the two critical lines at a given value of $\tilde{B}$,
$\Delta=(n^\star_c-n^\star_{\rm ld, c})/n^\star_c$. For $\Delta \lesssim
0.7$, the ratio $\Omega_{\rm max}$ is larger than 1.0 (corresponding
also to $\tilde{B}>1.0$), and the data can be fitted with a power law,
$\Omega_{\rm max} \propto \Delta^{-1.25}$, indicating a divergence
of this ratio towards $\tilde{B} \to \infty$. As a consequence, one
expects at least strong corrections to the asymptotic critical behavior
for large $\tilde{B}$ fields and, in the limit $\tilde{B}\to \infty$,
where the two critical points meet, the diffusion constant does not
vanish but becomes infinite.

\begin{figure}
\includegraphics[scale=0.3]{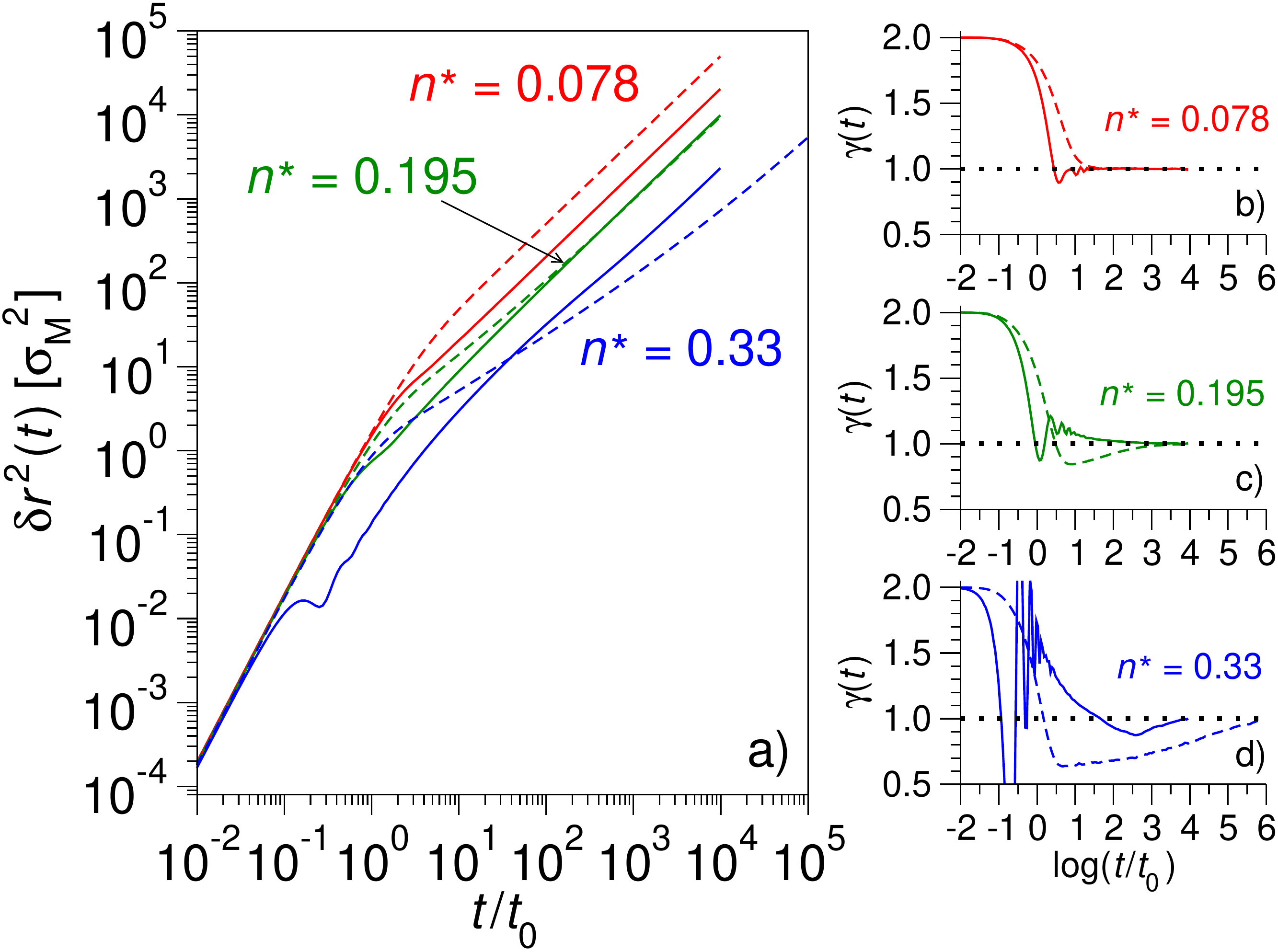}
\caption{\label{fig4} a) MSDs, $\delta r^2(t)$, for the densities
$n^\star = 0.078$, 0.195, and 0.33. The dashed lines are
for $\tilde{B}=0$, while the solid lines correspond to the
location of maximal diffusion at $\tilde{B}=0.36$, 1.02, and 7.09,
respectively. Panels b), c), and d) show the local exponents $\gamma(t)$,
corresponding to the MSDs in a).}
\end{figure}
On a microscopic scale, the qualitative change of the tracer particle
motion around $\tilde{B}=1$ can be analyzed in terms of MSDs. Figure
\ref{fig4}a displays MSDs at three different states of maximal diffusion,
$(n^\star_{\rm max}, \tilde{B})$ (cf.~the snapshots in Fig.~\ref{fig2}
at the same states). Also included are MSDs for $\tilde{B}=0$ (dashes
lines) at the corresponding densities. The onset of a directed motion
along the contour of the obstacles is associated with a superlinear
regime in the MSD at intermediate times for $t \gtrsim 3\,t_0$. This
is especially evident from the behavior of the local exponent of the
MSD, $\gamma(t)=d \log(\delta r^2(t)/d \log(t)$, which is shown in
Figs.~\ref{fig2}b)-d) for the three different densities.  At $n^\star =
0.078$ the diffusive regime is already reached around $t=10\,t_0$ and the
diffusion for $\tilde{B}=0.36$ is slightly slower than for $\tilde{B}=0$
due to the existence of the low-density critical point in the former
case. At $n^\star=0.195$, there is a superlinear regime for $1.0\,t_0
\lesssim t \lesssim 100\,t_0$ at $\tilde{B}=1.02$, while in the case
of $\tilde{B}=0$, a sublinear regime is seen in the same time range. A
similar effect, albeit much more pronounced, is present for $n^\star
= 0.33$. Here, the $\tilde{B}=0$ curve shows an extended sublinear
regime over about 2-3 orders of magnitude due to the vicinity of the
critical density $n^\star_{\rm c}$. This regime is almost suppressed
for $\tilde{B}=7.09$; instead a pronounced superlinear regime and a
faster transition towards normal diffusion is observed. This indicates
that particularly in a dense matrix the application of a magnetic field
$\tilde{B}\gg 1.0$ leads to a very efficient exploration of the matrix
due to the directed motion along the contour of the obstacle matrix.

{\it Summary and conclusions.} Using a combination of experiment and
simulation, we have studied the magneto-transport through two-dimensional
disordered Lorentz gases in the classical regime. Our focus was on the
non-monotonous behavior of the conductivity/diffusion which is observed
for magnetic fields $\tilde{B}\gtrsim 1.0$. These features cannot be
described by any of the existing kinetic theories. We emphasize that the
system under study is also related to the active motion of microswimmers
\cite{Zeitz2017} and thus has a more general relevance. The threshold
$\tilde{B}\approx 1.0$ marks the point where the cyclotron radius becomes
smaller than the obstacle radius. This leads to the change in the motion
of the tracer particle from a diffuse scatter by the obstacles to a
directed motion along the obstacle contour. The latter directed motion
is associated with an intermediate superlinear regime in the MSD that
becomes more pronounced with increasing $\tilde{B}$.

We have shown that one can draw a line of maximal diffusion into the
phase diagram that follows the law given by Eq.~(\ref{eq_bmax}). Along
this line, the diffusion constant ratio $\Omega_{\rm max}$ (see above)
diverges in the limit $\tilde{B} \to \infty$ (note that in this limit the
two critical lines intersect and thus $n^\star_{\rm c}=n^\star_{\rm ld,
c}$). Thus, in the limit $B\to \infty$ the directed motion dominates
the transport and leads to a divergence instead of a vanishing of
the diffusion coefficient. For finite $\tilde{B}$, at least strong
scaling corrections are expected in the vicinity of the two critical
points. Whether there is even a continuous change of the universality
class with respect to the two critical points with increasing $\tilde{B}$
is a subject of forthcoming studies.

%
{\it Achnowledgments.} We thank Thomas Franosch and Herbert Spohn for useful discussions. The
authors acknowledge financial support by  the German DFG, FOR 1394
(grant HO 2231/7-2). Computer time at the ZIM of the University of
D\"usseldorf is also gratefully acknowledged.
%


%

\end{document}